# Plagiarism Detection using ROUGE and WordNet

Chien-Ying Chen, Jen-Yuan Yeh, and Hao-Ren Ke

**Abstract**—With the arrival of digital era and Internet, the lack of information control provides an incentive for people to freely use any content available to them. Plagiarism occurs when users fail to credit the original owner for the content referred to, and such behavior leads to violation of intellectual property. Two main approaches to plagiarism detection are fingerprinting and term occurrence; however, one common weakness shared by both approaches, especially fingerprinting, is the incapability to detect modified text plagiarism. This study proposes adoption of ROUGE and WordNet to plagiarism detection. The former includes n-gram co-occurrence statistics, skip-bigram, and longest common subsequence (LCS), while the latter acts as a thesaurus and provides semantic information. N-gram co-occurrence statistics can detect verbatim copy and certain sentence modification, skip-bigram and LCS are immune from text modification such as simple addition or deletion of words, and WordNet may handle the problem of word substitution.

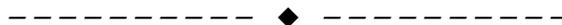

## 1 INTRODUCTION

PLAGIARISM is the use of other people's work/idea as one's own without crediting the original owner.

Although there are different plagiarism detection approaches, each method has its pros and cons. One common weakness is the vulnerability to text modification which can be achieved through addition, deletion and substitution of words, and also change of sentence structure or word order. This study proposes a system based on ROUGE [1] and WordNet [2] to conquer most of the text alteration strategies. ROUGE calculates n-gram co-occurrence statistics to evaluate the quality of a candidate summary with regard to one or more reference summaries. The basic concept underlying ROUGE is that the more similar a candidate summary is to a reference summary, the better the qualities of summary. The same concept can be applied to plagiarism detection – the more similar a candidate text is to a reference text, the more likely that plagiarism has occurred. Moreover, two methods in ROUGE, longest common subsequence (LCS) and skip-bigram, may work on certain types of plagiarism, i.e. the addition or deletion of words.

WordNet is a dictionary-like database developed by Cognitive Science Laboratory of Princeton University [2]. As people can substitute words in the original text with synonyms when they plagiarize, n-grams that only consider exact match is unable to detect the substitutions, resulting false negative between the two sentences. Therefore, WordNet may be helpful when analyzing a sentence pair for this type of plagiarism, because it can be applied to find implicit relationship between two words.

The rest of this paper is organized as follows. Related works are described in Section 2, methodology in Section 3, experiments and discussion in Section 4, and conclusion in Section 5.

## 2 RELATED WORKS

A considerable amount of research has focused on plagiarism detection. Fig. 1 provides an overview about the development of plagiarism detection. The classification is derived from the taxonomy in [3]. As it indicates, the plagiarism detection methods can be categorized into three main categories: *fingerprinting*, *term occurrence*, and *style analysis*.

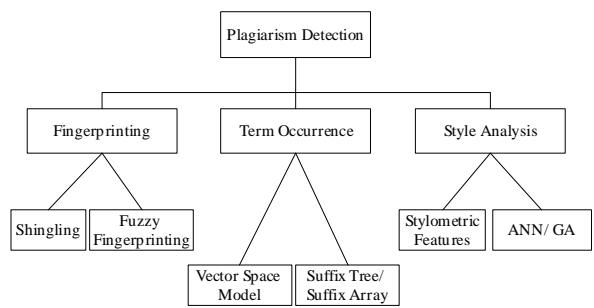

Fig. 1 Classification of Plagiarism Detection Methods

- *Chien-Ying Chen is with the Institute of Information and Management, National Chiao-Tung University, Taiwan, R.O.C. E-mail: chen_chien_ying@hotmail.com.*
- *Jen-Yuan Yeh is with the Department of Computer and Information Science, National Chiao-Tung University, Taiwan, R.O.C. E-mail: jenyuan.yeh@gmail.com.*
- *Hao-Ren Ke (Corresponding Author) is with the Graduate Institute of Library and Information Studies, National Taiwan Normal University, Taiwan, R.O.C. E-mail: clavenke@ntnu.edu.tw.*

## 2.1 Fingerprinting

Fingerprinting can be considered as the most widely adopted approach in plagiarism detection. Manber [4] aimed to find out similar documents in a database by Rabin fingerprint scheme [5], which was applied to generate a unique identity (fingerprint) for each document. Rabin fingerprint scheme or hash function as often used interchangeably, can transform a sequence of substring into an integer. A good scheme/function should generate the same integer for the same substring; on the other hand, it should generate different integer for each unique substring to ensure consistency and avoid undesirable collisions of fingerprints. Four factors are need to be considered when generating fingerprints, including substring selection, substring number (fingerprint resolution), substring size (fingerprint granularity), and substring encoding ([3], [6]).

Fingerprinting was first applied to the field of plagiarism detection in COPS [7] for copy detection. In COPS, the smallest detection unit was a sentence, but multiple sentences could form a larger unit of detection called chunk. COPS included a document database to save newly processed documents and compare a suspect document with registered documents. Later, SCAM (Stanford Copy Analysis Mechanism) [8] was developed based on COPS. Nevertheless, SCAM focused on word-based overlap as fingerprints were generated in unit of word instead of sentence. The change led to better effectiveness in detecting partial copy but more false positives as a tradeoff. Majority of later approaches focus on various aspects of fingerprinting; usually, different strategies are employed or other techniques are integrated with fingerprinting. Variations include shingling and fuzzy fingerprinting. The former is a combination of fingerprinting and unique n-gram substrings [9], while the latter adopts inverse document frequency (idfs) into substring selection to pick out feature terms [10].

## 2.2 Term Occurrence

Term occurrence is probably the most intuitive approach because lexical words contain explicit information on the text and they can be analyzed to determine the similarity between two documents. One assumption is that the more terms both documents have in common, the more similar they are. Term occurrence has been applied to a range of studies such as automatic evaluation of summaries, automatic evaluation of machine translation, and common information retrieval (IR) problems like clustering and categorization. Due to a common purpose between the aforementioned studies and plagiarism detection, i.e. determining similarities between documents, application of term occurrence in plagiarism detection seems promising.

CHECK [11], which incorporates a well-known IR model - vector space model (VSM), is a plagiarism detection method that first parses a document into a tree structure before comparing two documents. The root node contains the overall information of a document while the internal and leaf nodes contain information of sections and paragraphs respectively. The authors called the tree structure the document tree and the information within as structural characteristics (SC) of the document. Each node in the tree is expressed as a weighted vector. If the similarity measure (for example, cosine similarity) between two root nodes exceeds a certain threshold, the two corresponding documents are thought to be similar in content and child nodes will be compared. The similarity between child nodes is compared in the same manner as the root nodes, and the above comparison process are performed recursively until the similarity measure falls below the threshold or when leaf nodes are reached.

Zaslavsky, Bia, and Monostori [12] utilized suffix trees, each of which contains all the suffixes of a string and therefore all the substrings as well. When applied at document level, the suffix tree method, together with the matching statistics algorithm, is able to find overlapping chunks between two documents. Khmelev & Teahan [13], instead of using suffix trees, adopted the idea of suffix arrays to reduce the memory problem found in suffix trees.

## 2.3 Style Analysis

Style analysis is the most special approach to plagiarism detection, because unlike other methods, it requires no reference corpus and it focuses more on implicit information than explicit information on the texts. The basic principles behind style analysis are that every author has his/her own writing style, that each person's writing style should remain consistent throughout the text, and that the characteristics of a style is hard to manipulate or imitate, making the plagiarized portion of work to stand out in the text implicitly [14]. The writing style of an author includes richness in vocabulary, sentence lengths, and the number of closed class words and open class words used. Analysis of those measures enables researchers to turn the abstract idea of writing style into realistic numbers [3]. Although style analysis may not need a reference corpus, it needs to be trained to learn about rules of one's writing. Hence, various artificial neural networks (ANNs) and genetic algorithms (GAs) have been applied to analyze style and authorship [14]. The trained ANNs or GAs will be able to recognize the style of a particular author and therefore articles written by the author.

## 3 METHODOLOGY

Having discussed about existing plagiarism detection methods, the methods proposed in this research will be discussed next.

Before two documents can be compared with each other, they have to be preprocessed. There are a total of six preprocessing steps, including tokenization, punctuation removal, lowercase conversion, part-of-speech (POS) tagging, stopword removal, and Porter stemming [15].

The proposed system employs four methods, n-gram analysis up to 4-gram, LCS, skip-bigram, and WordNet, for detecting plagiarism. The former three methods are ROUGE-based. The details of these methods will be discussed in the following subsections.

## 3.1 N-grams

Each token (word) in a sentence is a unigram. Before comparing the sentences, every unique unigram and its number of occurrence(s) in the sentence will be recorded. Beginning with the first sentence of the reference document, every unique unigram is compared with all unique unigrams in every sentence of the candidate document, followed by the second sentence of the reference document and so on and so forth. Overall, all reference sentences will be compared with all candidate sentences for a total of M x N times, where M and N are the number of sentences in the reference and candidate documents respectively.

---

Candidate Sentence:
*the the the the the the the*

Reference Sentence:
*the cat is on the mat*

Standard Precision: 7/7
Clipped Precision: 2/7

---

Fig. 2 Example of Clipped Precision

The number of overlapping unigram(s) between two sentences, $S_u^R$ from the reference document and $S_v^C$ from the candidate document, will be counted. The overlapping total, i.e. numerator of (1) and (2), is divided by the length of the reference sentence and length of the candidate sentence respectively in order to calculate *recall* and *precision*. The numerator is the smaller number of occurrence of the overlapping unigrams in the two sentences. This is to avoid false positive in certain cases, in which a particular unigram are found in both the candidate and reference sentences but appears more in the candidate sentence. Such modification is called *clipping* [16]. Fig. 2 is an example from BLEU [16], in which the word, *the*, appears in both the reference and candidate sentences two times and seven times respectively. If according to (1) without the clipping mechanism, the precision score contributed by this word will be 7/7, which is clearly exaggerated. However, if the score is clipped it becomes 2/7, which is more reasonable.

$$R - N(S_u^R, S_v^C) = \frac{\sum\limits_{\text{n-gram} \in S_u^R \text{ and } S_v^C} \text{Count}_{\text{clip}} (\text{n-gram})}{\sum\limits_{\text{n-gram} \in S_u^R} \text{Count (n-gram')}} \quad (1)$$

$$P - N(S_u^R, S_v^C) = \frac{\sum\limits_{\text{n-gram} \in S_u^R \text{ and } S_v^C} \text{Count}_{\text{clip}} (\text{n-gram})}{\sum\limits_{\text{n-gram} \in S_v^C} \text{Count (n-gram')}} \quad (2)$$

N-gram (including unigram) score is expressed as (3).

$$Score - N(S_u^R, S_v^C) = \frac{2*(R-N)*(P-N)}{(R-N)+(P-N)} \quad (3)$$

## 3.2 Longest Common Subsequence (LCS)

LCS is the longest in-sequence string of matched tokens between two sentences ([1], [17]). In unigram matching,

the position of a matched token is not a constraint. As long as a unigram co-occurs in both sentences, it will contribute to the similarity between two sentences. Although LCS is also based on matching unigrams, it only considers matched tokens that form the longest in-sequence subsequence of the reference sentence. In other words, even if a unique unigram is in both the reference and candidate sentences, but if it is out of order with other matched tokens, it is not included in the LCS and will not contribute to the LCS score; furthermore, if more than one common subsequence exists, LCS will only reflect the longest subsequence among them. Another characteristic of LCS is that unlike n-grams (excluding unigram), LCS allows skip of matched tokens, which need not be strictly consecutive. Fig. 3, taken from ROUGE [1], illustrates how LCS is derived.

---

Candidate sentence 1: Police kill the gunman

Candidate sentence 2: The gunman kill police

Reference sentence: Police killed the gunman

The LCS between Reference sentence and Candidate sentence 1 is *police the gunman* while the LCS between Reference sentence and Candidate sentence 2 is *the gunman, excluding police*. The first pair of sentences shows the skipping nature of LCS and the second pair of sentences shows the in-sequence rule that bounds LCS.

---

Fig. 3 Example of LCS

LCS score can be expressed as (6), which is derived from (4) and (5).

$$R - LCS(S_u^R, S_v^C) = \frac{LCS(S_u^R, S_v^C)}{\sum\limits_{\text{unigram} \in S_u^R} \text{Count (unigram)}} \quad (4)$$

$$P - LCS(S_u^R, S_v^C) = \frac{LCS(S_u^R, S_v^C)}{\sum\limits_{\text{unigram} \in S_v^C} \text{Count (unigram)}} \quad (5)$$

$$Score - LCS(S_u^R, S_v^C) = \frac{2*(R-LCS)*(P-LCS)}{(R-LCS)+(P-LCS)} \quad (6)$$

## 3.3 Skip-Bigram

Skip-bigram is a variation of bigram. The difference between them lies in the formation of bigrams. For skip-bigram, bigrams are formed not only by consecutive tokens, but also by other in-sequence tokens. Skip distance, *d*, has to be set before counting skip-bigrams in a sentence. Skip distance is the maximum number of tokens in between any two combining tokens. After the skip distance is determined, all the skip-bigrams within a sentence can be found. Let $w_1, w_2, ..., w_n$ be a sentence. Starting with $w_1$, it will form skip-bigrams with the fol-lowing $d+1$ words, followed by $w_2$, which forms skip-bigrams with another in-sequence $d+1$ words. The process stops when $w_{n-1}$ forms the last skip-bigram with $w_n$. Fig. 4 shows how skip-bigrams of a four-word string are formed. After finding the skip-bigrams for the sentence pair, the same steps for n-grams can be used to compare the skip-bigrams, and the skip-bigram score is expressed in (7)-(9), where the lengths of $S_u^R$ and $S_v^C$ are $p$ and $q$, respectively.

---

For a given sequence: andy eats an apple

When d=2, skip-bigrams generated will be as follows: Start with the first word *andy*, it can form a bigram with the furthest token, *apple*, and followed by *eats* and *an* respectively. When *andy* has formed bigrams with all possible tokens, *eats* will form bigrams with *an* and *apple*. Finally, the last skip-bigram is *an apple*.

There are a total of six skip-bigrams by the sequence above when d=2. They are as shown:
*andy eats, andy an, andy apple, eats an, eats apple, an apple*
*andy eats, andy an, andy apple, eats an, eats apple, an apple*

---

Fig. 4 Example of Skip-bigram Formation

$$R-Skip(S_u^R, S_v^C) = \frac{SKIP2_{clip}(S_u^R, S_v^C)}{(p-d-1)(d+1)+\sum_{r=0}^{d-1}d-r} \quad (7)$$

$$P-Skip(S_u^R, S_v^C) = \frac{SKIP2_{clip}(S_u^R, S_v^C)}{(q-d-1)(d+1)+\sum_{r=0}^{d-1}d-r} \quad (8)$$

$$Score-Skip(S_u^R, S_v^C)_{\substack{1\le u\le M \\ 1\le v\le N}} = \frac{2*(R-Skip)*(P-Skip)}{(R-Skip)+(P-Skip)} \quad (9)$$

## 3.4 WordNet

WordNet is a large lexical database of English. WordNet groups nouns, verbs, adjectives and adverbs into sets of cognitive synonyms (called *synsets*), each indicating a distinct concept. Synsets are interlinked by means of conceptual-semantic and lexical relations [2].

In this study, WordNet is integrated for discovering synonyms and semantic-related words (unigrams). In the unigram method, no score is given if two comparing unigrams are not exactly the same. However, extra steps are taken after integrating WordNet. Following the same procedures in Section 3.1, a unique unigram from the reference sentence will be matched against all the unique unigrams in the candidate sentence. However, if there is not an exact match, the relationships between the words in WordNet will be looked up.

Besides the word itself, the POS of the word is necessary as well to retrieve its information in WordNet. If both words with their specific POS can be found in the WordNet, their lexical and semantic relationship can be determined. If not, no further action is taken and this pair of words is considered irrelevant.

Two different measures are considered to determine the relationship between two words. They will be discussed as follows.

### 3.4.1 Synonyms-based measure

The similarity between two synsets is measured by Jaccard's coefficient, as shown in (10), where $s_k^i$ is synset $k$ of unigram $i$ while $s_l^j$ is synset $l$ of unigram $j$. The numerator is the overlapping synonyms between synsets $k$ and $l$ while the denominator is the distinct union of synonyms of synsets $k$ and $l$. Fig. 5 is an example of how two synsets are matched:

$$sim(s_k^i, s_l^j) \quad \frac{|k \cap l|}{|k \cup l|} \quad (10)$$

---

Synsets of *shouts (verb)*:
1. shout
2. shout, shout out, cry, call, yell, scream, holler, hollo, squall
3. exclaim, cry, cry out, outcry, call out, shout
4. abuse, clapperclaw, blackguard, shout

Synsets of *yells (verb)*:
1. shout, shout out, cry, call, yell, scream, holler, hollo, squall
2. yell, scream

As shown above, the verb, *shouts*, has four synsets and the verb, *yells*, has two synsets in WordNet. By looking at synsets 1 of both words, there is only one overlapping synonym while the distinct union of synonyms is 9. Therefore, the Jaccard's coefficient for these two synsets will be 1/9.

---

Fig. 5 Example of Jaccard's Coefficient between Two Synsets

Each of the synsets of a unigram will be compared with all the synsets of the other unigram. The highest score will be used as the similarity score between two unigrams, as expressed in (11), where $W_i^R$ is unigram $i$ in the reference sentence with $m$ synsets and $W_j^C$ is unigram $j$ in the candidate sentence with $n$ synsets

$$sim(w_i^R, w_j^C) = Arg \max_{\substack{1\le k\le m \\ 1\le l\le q}} sim(s_k^i, s_l^j) \quad (11)$$

The same steps are repeated until a reference unigram $W_i^R$ has compared with all the unique unigrams in the candidate sentence. The highest similarity score among the calculated scores will be chosen as the similarity score between word $W_i^R$ and a given sentence $S_v^C$, as shown in (12), where $W_j^C$ is a unigram in $S_v^C$.

$$sim(w_i^R, S_v^C) = Arg \max_{1\le j\le q} sim(w_i^R, w_j^C) \quad (12)$$

Finally, the plagiarism score for any pair of reference sentence and candidate sentence can be expressed as (13)-(15).

$$R-Syn(S_u^R, S_v^C) = \frac{\left(\sum_{i=1}^{p} sim_{clip}(w_i^R, S_v^C)\right)}{\sum_{unigram \ \in S_u^R} Count \ (unigram)} \quad (13)$$
$$\scriptstyle 1 \le u \le M \atop 1 \le v \le N$$

$$P-Syn(S_u^R, S_v^C) = \frac{\left(\sum_{i=1}^{p} sim_{clip}(w_i^R, S_v^C)\right)}{\sum_{unigram \ \in S_v^C} Count \ (unigram)} \quad (14)$$
$$\scriptstyle 1 \le u \le y \atop 1 \le v \le z$$

$$Score-Syn(S_u^R, S_v^C) = \frac{2*(R-Syn)*(P-Syn)}{(R-Syn)+(P-Syn)} \quad (15)$$
$$\scriptstyle 1 \le u \le M \atop 1 \le v \le N$$

### 3.4.2 Relationship-based measure

Synsets of *shouts (verb)*:
1. shout
2. shout, shout out, cry, call, yell, scream, holler, hollo, squall
3. exclaim, cry, cry out, outcry, call out, shout
4. abuse, clapperclaw, blackguard, shout

Synsets of *yells (verb)*:
1. shout, shout out, cry, call, yell, scream, holler, hollo, squall
2. yell, scream

Although there are a total of eight possible combinations of synsets between the two verbs, *shouts* and *yells*, these two words are linked by two pairs of synsets. Synset 1 of *shouts* and synset 2 of *yells*; synset 2 of *shouts* and synset 1 of *yells*. The first pair only has hypernym relationship of depth 1, while the second pair has both hypernym and hyponym relationships of depth 0. Since depth 0 is the closest relationship possible, the relationship between *shouts* and *yells* is represented by the second pair.

Fig. 6 Example of Hypernym/Hyponym Relationship between Two Words

Relationship here refers to hypernym and hyponym relationships in WordNet. Same steps in synonym-based method are carried out to obtain the *senses*. After that, hypernym and hyponym relationships between two senses can be found. The term senses are used here instead of synsets because synonyms are not the focus but how each sense/meaning of a word is semantically related to other senses of the other word. Again, all senses of the reference unigram have to be compared with all the other senses of the candidate unigram. The relationship is expressed hierarchically in terms of depth. If two words are actually in the same synset, the depth is zero. This research only considers relationship depth within three levels in the hierarchy. Fig. 6 is an example of how hypernym/hyponym relationships between two words are determined.

The calculation of the relationship-based score is similar to that of the synonym-based score, and is expressed in (16)-(20).

$$RS(w_i^R, w_j^C) = Arg \max(Arg \max_{\substack{1 \le k \le m \\ 1 \le l \le n}} hypernym(s_k^i, s_l^j),$$
$$Arg \max hyponym(s_k^i, s_l^j)) \quad (16)$$

$$RS(w_i^R, S_v^C) = Arg \max_{1 \le j \le q} RS(w_i^R, w_j^C) \quad (17)$$

$$R-RS(S_u^R, S_v^C) = \frac{\left(\sum_{i=1}^{p} sim_{clip}(w_i^R, S_v^C) \Box wt\right)}{\sum_{unigram \ \in S_u^R} Count \ (unigram)} \quad (18)$$
$$\scriptstyle 1 \le u \le M \atop 1 \le v \le N$$

$$P-RS(S_u^R, S_v^C) = \frac{\left(\sum_{i=1}^{p} sim_{clip}(w_i^R, S_v^C) \Box wt\right)}{\sum_{unigram \ \in S_v^C} Count \ (unigram)} \quad (19)$$
$$\scriptstyle 1 \le u \le y \atop 1 \le v \le z$$

$$Score-RS(S_u^R, S_v^C) = \frac{2*(R-RS)*(P-RS)}{(R-RS)+(P-RS)} \quad (20)$$

*wt* in (18) and (19) denotes the weight for hypernym/hyponym relationships. Through empirical experiments, assignment of weight for each depth will be as follows: 1.0 for depth 0, 0.85 for depth 1, 0.5 for depth 2, and 0.2 for depth 3.

## 4 EXPERIMENTS AND DISCUSSION

### 4.1 Data Sets

There are two manually generated data sets for evaluating the proposed methods. One of the sets contains 978 pairs of sentences while the other contains 100 pairs of sentences. These two sets will be referred to as the *abstract* set and the *paraphrased* set respectively hereafter. The abstract set was based on the observation that abstracts of some papers are actually taken from parts of the main text. Such characteristic may be utilized to simulate the plagiarism scenario by treating the abstract as the candidate of plagiarism and the main text as the source (reference) being plagiarized. In view of this, a collection of research papers with similar theme were retrieved from research databases. Each abstract sentence was compared with the main text using six different methods namely n-grams (unigram to 4-gram), LCS and skip-bigram. Top five matching in each method were recorded regardless of the scores. In other words, each abstract sentence produced 30 pairs of sentences with the main text; however, if there were repeated pairs within the 30 pairs, only one of them would be included in the abstract corpus. There were 1000 unique pairs of sentences out of 19 research papers in the end. In order to ensure the validity of the corpus, four people who understood the concept of plagiarism were asked to annotate the abstract corpus. Each person was given 500 pairs in a manner that each pair would be annotated twice. After the annotation was completed, kappa statistics [18] was applied to ensure the reliability and validity of the annotation by measuring the agreement between the annotators. In the end, the kappa score was 0.791, which fell under the category of Substantial.

Sentence pairs which were annotated differently by the annotators were removed, and the end product was 978 pairs of sentences in which only 32 pairs were annotated as candidates of plagiarism. It is observed that most of the 32 pairs were verbatim type of plagiarism, and there were not as many substitutions of words.

The paraphrased set was generated by retrieving short plagiarism examples of a few sentences long from the Internet. Hence, the query "plagiarism examples" was sent to Google, and only paraphrased plagiarism examples were retrieved. Paraphrased set consists of plagiarism types like addition, deletion or substitution of words in the original content, change of sentence structure, and partial verbatim copy. A total of 100 plagiarism examples were retrieved. The plagiarizing sentences were paired up with the corresponding original sentences manually; therefore each pair was a valid example of plagiarism.

## 4.2 Experiment Settings

To test the effectiveness of the proposed methods in detecting plagiarism, the experiments calculated the score of each method, i.e. (3), (6), (9), (15), and (20) with different thresholds. The values of the abovementioned scores range from 0.0 to 1.0. During the experiments, each reference sentence only compared with its corresponding candidate sentence, not with all the candidate sentences. By comparing the score of the sentence pair with the threshold, there were four different possible outcomes namely true positive (TP), false positive (FP), true negative (TN), and false negative (FN). If the score was larger than the threshold and the pair was annotated as *plagiarism*, a TP would be recorded; on the other hand, a FP would be recorded instead if the pair was annotated otherwise. The same logic applied when judging TN and FN but with opposite criteria, i.e. the score was smaller than the threshold and the pair was annotated as *not plagiarism*.

By looking at the problem from an information retrieval perspective, the number of plagiarism pairs was the number of relevant documents. Therefore the performance of each method could be evaluated in terms of recall, precision, and F-measure [15].

Each method was tested under several specific preprocessing settings. For ROUGE-based methods, there are four possible settings: both stopwords and stemming are applied (SW+SM), stopwords are removed (SW), stemming is applied (SM), and neither stopwords nor stemming are applied (No Pre). Meanwhile, there are only two possible settings for The WordNet-based methods: SW and no preprocessing. Stemming is not compatible with WordNet because some words have stems that WordNet cannot recognize. For example, "happy" will be stemmed to "happi"; in such case, WordNet is unable to find a match in its database.

## 4.3 Results and Discussion

### 4.3.1 Recommended Settings of the ROUGE-based Methods for Abstract Set

To recommend a setting for each method, comparison between the performances of the same method under different settings was necessary. The following describes the principles for choosing the setting for the unigram method, and the same principle can be used for determining the settings for all other ROUGE-based methods. It can be seen from Fig. 9 that F-measure with No Pre - F(No Pre) and F-measure with SM - F(SM) have the two highest F-measures at threshold=0.5, respectively, while F(SW+SM) and F(SW) formed smoother curves on the graph. By only looking at threshold≥0.5, F(SW+SM) and F(SW) obviously performed better than F(SM) and F(No Pre); hence the choices were cut down to two: F(SW+SM) and F(SW).

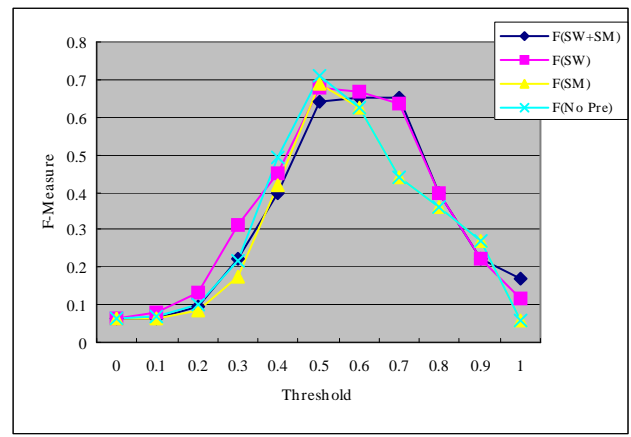

Fig. 7 Performances of Unigram under Different Preprocessing Settings

Here, the information of TPs, FPs, TNs, and FNs could provide some insights from another perspective. As shown in TABLE 1 and TABLE 2, at threshold=0.5, the number of FPs for both settings were unacceptable, but FPs dropped significantly at threshold=0.6. And by observing threshold≥0.6, F(SW) had slightly lower number of FPs and TPs than F(SW+SM). Up to this point, a recommended setting for the unigram method was determined – stopwords removal with threshold=0.6. One reason for sacrificing TPs in exchange of lower FPs is that the definition of plagiarism does not depend on the number of plagiarism instances, but depend on whether or not a plagiarism instance really exists. With this criterion in mind, as long as the number of TPs is substantial, lower FPs will be the top priority. The other reason is that no system so far can guarantee fully automatic detection, human judgment is mandatory at the very end of the detection process; therefore by lowering the number of FPs, less time is required to filter out FPs by the evaluator.

TABLE 1 Results of F(SW+SM)

| Threshold | TP | FP | TN | FN |
|---|---|---|---|---|
| 0.5 | 23 | 17 | 929 | 9 |
| 0.6 | 17 | 3 | 943 | 15 |
| 0.7 | 16 | 1 | 945 | 16 |
| 0.8 | 8 | 0 | 946 | 24 |
| 0.9 | 4 | 0 | 946 | 28 |
| 1.0 | 3 | 0 | 946 | 29 |

TABLE 2 2 Results of F(SW)

| Threshold | TP | FP | TN | FN |
|---|---|---|---|---|
| 0.5 | 22 | 11 | 935 | 10 |
| 0.6 | 17 | 2 | 944 | 15 |
| 0.7 | 15 | 0 | 946 | 17 |
| 0.8 | 8 | 0 | 946 | 24 |
| 0.9 | 4 | 0 | 946 | 28 |
| 1.0 | 2 | 0 | 946 | 30 |

TABLE 3 shows the recommended settings for the remaining ROUGE methods, which were determined under the same principle.

TABLE 3 Recommended Settings for ROUGE-based Methods

| Methods | Recommended Setting |
|---|---|
| Unigram | Stopwords, Threshold=0.6 |
| Bigram | No Preprocessing, Threshold=0.4 |
| Trigram | Stemming, Threshold=0.3 |
| 4-gram | Stopwords, Threshold=0.3 |
| Skip-bigram | Stopwords & Stemming, Threshold=0.3 |
| LCS | Stopwords & Stemming, Threshold=0.5 |

### 4.3.2 Recommended Settings of the WordNet-based Methods for Abstract Set

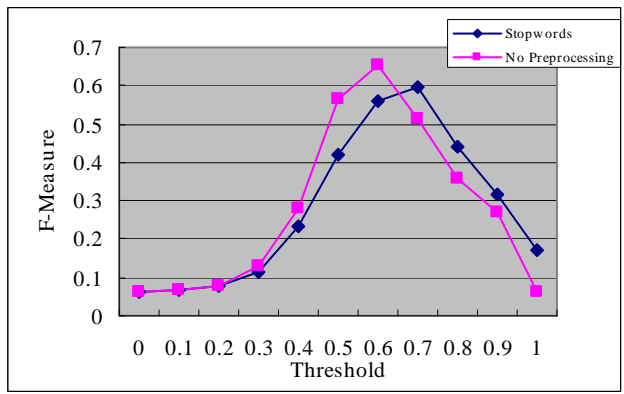

Fig. 8 Performances of the Relationship-based Method under Different Settings

The performances of the relationship-based method under different settings are shown in Fig. 8, which shows that SW performed better than No Pre at threshold=0.7 and beyond; therefore, the recommended setting for relationship-based method was with stopwords removed and threshold=0.7.

Recommendation of the setting for synonyms-based method can be indicated by Fig. 9. Synonyms-based method performed better with stopwords removed and the number of FPs at threshold=0.6 were less than the number of FPs at threshold=0.5. Therefore, synonyms-based method should be applied at threshold=0.6 with stopwords removed.

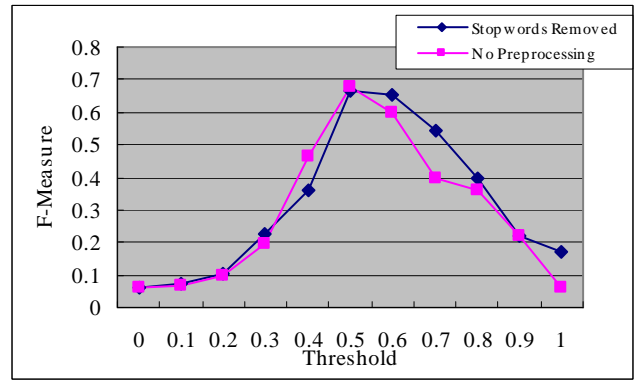

Fig. 9 Performances of the Synonyms-based Method under Different Settings

### 4.3.3 Discussion of the WordNet-based Methods for Abstract Set

Because both synonyms-based and relationship-based methods were derived from unigram, unigram was the best candidate among ROUGE-based methods to be compared with the WordNet-based methods. However, to ensure valid comparison between the methods, like the WordNet-based methods, POS tags had to be included in unigram during the matching of tokens. In addition to this modification, unigram (SW+SM) was compared with the two WordNet-based methods (SW) – Fig. 10, and unigram (SM) was compared with the other two methods (No Pre) – Fig. 11. The main reason for such match-ups was due to the fact that WordNet transformed the words into their original form in the database. For example, from "paid" to "pay". Such process was similar to the concept of stemming and hopefully the match-ups would make the comparisons more meaningful.

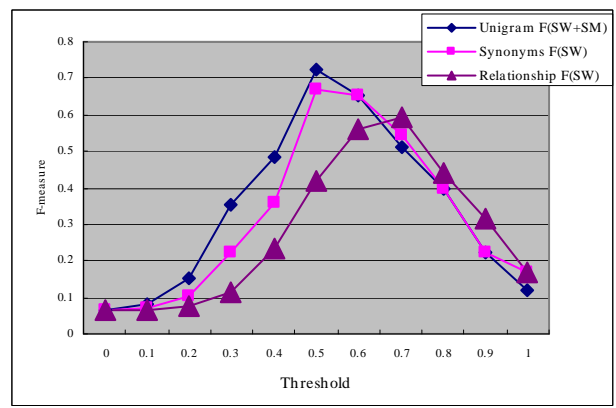

Fig. 10 F-Measures Comparison with Stopwords Removed for the WordNet-based Methods and Unigram

Fig. 10 and Fig. 11 showed that on overall it was hard to tell if the WordNet-based methods were better than unigram in the abstract set. However, by looking at thresholds≥0.6, i.e. beyond the recommended thresholds of the three methods, the WordNet-based methods did perform better than unigram. The outcomes were probably due to the nature of the abstract set – most valid plagiarism pairs in the corpus were made up of verbatim type of plagiarism, and there were not as many substitutions of words. As a result, WordNet was not being fully utilized and could not be of much help.

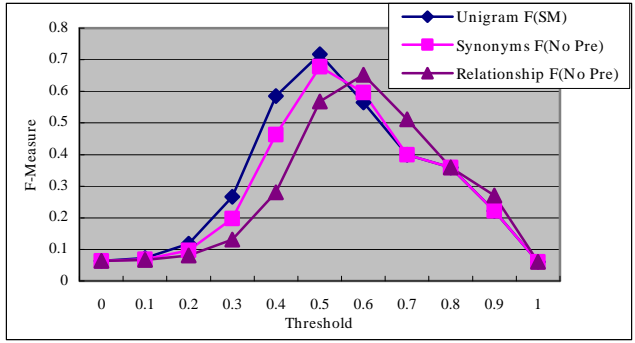

Fig. 11 F-Measures Comparison with No Preprocessing for the WordNet-based Methods and Unigram

### 4.3.4 Discussion of the WordNet-based Methods for Paraphrased Set

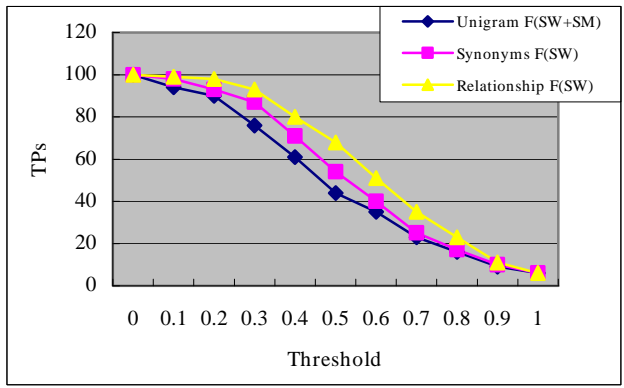

Fig. 12 TPs of Unigram and WordNet-based Methods with Stop-words Removed

In the previous section, the WordNet-based methods performed moderately in the *abstract* set. Further evaluation on the WordNet-based methods was done with the *paraphrased* set. Evaluation was based on the number of TPs for each method and comparison of the results was made. Again, POS tags were included in unigram and the same match-ups for preprocessing were deployed. As the nature of plagiarism examples changed, the results were very different from Section 4.3.3. Fig. 12 and Fig. 13 show the same pattern of results, with relationship-based method on top, followed by synonyms, and unigram was at the bottom of the graph. Although this evaluation could not determine the effectiveness of the WordNet-based methods in identifying TNs, one affirmation was that the WordNet-based methods were able to identify substitution of words better than unigram since this was the only difference between these two approaches.

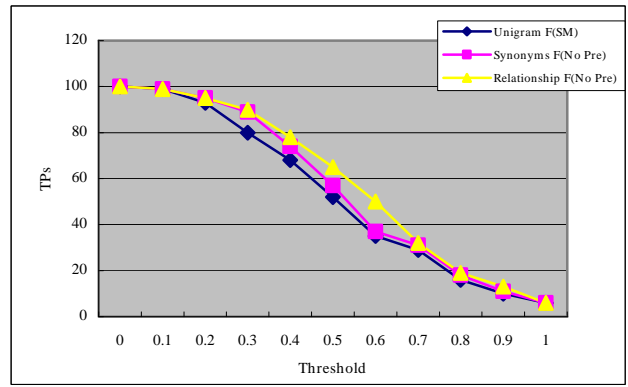

Fig. 13 TPs of Unigram and WordNet-based Methods with No Pre-processing

### 4.3.6 Strengths and Weaknesses of Each Method

At this point, ideal threshold and preprocessing setting for each method were empirically determined. TABLE 4 shows the performance of all the methods under different settings at their recommended threshold. By observing TABLE 4, one interesting discovery was made. Under every different setting, the highest F-measure was either from Skip-bigram or LCS. One possible explanation may be that both allow gaps between matching tokens but at the same time require tokens to be in-sequence. These two rules balance the values of recall and precision, and lead to higher F-measure.

TABLE 4 F-measures under Different Settings Recommended Threshold for Abstract Set

|  | SW+SM | SW | SM | No Pre |
|---|---|---|---|---|
| Unigram, T=0.6 | 0.65 | 0.67 | 0.63 | 0.63 |
| Bigram, T=0.4 | 0.65 | 0.67 | 0.67 | 0.67 |
| Trigram, T=0.3 | 0.57 | 0.58 | 0.63 | 0.63 |
| 4-gram, T=0.3 | 0.57 | 0.57 | 0.57 | 0.57 |
| Skip-bigram, T=0.3 | 0.63 | 0.63 | 0.69 | 0.67 |
| LCS, T=0.5 | 0.70 | 0.70 | 0.65 | 0.63 |
| Synonyms-based, T=0.6 | NA | 0.65 | NA | 0.60 |
| Relationship-based, T=0.7 | NA | 0.60 | NA | 0.51 |

To assess the specific strengths and weaknesses of the methods, a few examples were picked from the *paraphrased* set and tested with all the methods. The results were analyzed and conclusions were drawn from the analysis. The scores in this section refer to the scores between two sentences under different methods in Section 3.

Example 1:

Candidate Sentence: brown dwarfs **are difficult to locate and** rank among the most elusive objects in the universe
Reference Sentence: brown dwarfs rank among the most elusive objects in the universe

For example 1, all the methods had relatively high scores (TABLE 5). This indicates that if consecutive new words are inserted into the reference sentence and the lengths of the sentences do not differ much, all the methods still should be able to detect plagiarism. The zeros from 4-gram were due to the fact that there was not any matching four gram after stopwords were removed. NAs in TABLE 5 and the following tables mean that synonyms-based and relationship-based methods were not tested under those settings as mentioned in Subsection 4.3.3.

TABLE 5 Scores of Example 1

| Methods | SW+SM | SW | SM | No Pre |
|---|---|---|---|---|
| Unigram | 0.86 | 0.86 | 0.81 | 0.81 |
| Bigram | 0.67 | 0.67 | 0.72 | 0.72 |
| Trigram | 0.40 | 0.40 | 0.61 | 0.61 |
| 4-gram | 0.00 | 0.00 | 0.57 | 0.57 |
| Skip-bigram | 0.40 | 0.40 | 0.60 | 0.60 |
| LCS | 0.86 | 0.86 | 0.81 | 0.81 |
| Synonyms | NA | 0.79 | NA | 0.78 |
| Relationship | NA | 0.71 | NA | 0.74 |
| Unigram (POS) | 0.71 | 0.71 | 0.74 | 0.74 |

Example 2:

Candidate Sentence: in this view meaning is determined by **the** real world and is **therefore** external to the **learner**

Reference Sentence: meaning is **eventually** determined by **this** real world and is **thus** external to the **understander**

TABLE 6 Scores of Example 2

| Methods | SW+SM | SW | SM | No Pre |
|---|---|---|---|---|
| Unigram | 0.71 | 0.71 | 0.75 | 0.75 |
| Bigram | 0.50 | 0.50 | 0.47 | 0.47 |
| Trigram | 0.40 | 0.40 | 0.21 | 0.21 |
| 4-gram | 0.25 | 0.25 | 0.08 | 0.08 |
| Skip-bigram | 0.53 | 0.53 | 0.50 | 0.50 |
| LCS | 0.71 | 0.71 | 0.68 | 0.68 |
| Synonyms | NA | 0.71 | NA | 0.75 |
| Relationship | NA | 0.71 | NA | 0.75 |
| Unigram (POS) | 0.71 | 0.71 | 0.75 | 0.75 |

As indicated in TABLE 6, in Example 2, bigram, trigram, and 4-gram had lower scores than they had in Example 1 because of addition, deletion, and substitution of words throughout the candidate sentence, causing less matching of consecutive tokens. On the other hand, unlike the three methods above, skip-bigram and LCS allowed in-sequence skip within the sentence; as a result, skip-bigram was slightly better than bigram, and LCS still had satisfactory F-measures.

Example 3:

Candidate Sentence: those complexes that contain paired electrons are **repelled** by a magnetic field and are said to be **diamagnetic** whereas those with no paired electrons are attracted to such a field and are called **paramagnetic**

Reference Sentence: those complexes that contain unpaired electrons are **attracted** into a magnetic field and are said to be **paramagnetic** while those with no unpaired electrons are repelled by such a field and are called **diamagnetic**

TABLE 7 Scores of Example 3

| Methods | SW+SM | SW | SM | No Pre |
|---|---|---|---|---|
| Unigram | 0.85 | 0.85 | 0.88 | 0.88 |
| Bigram | 0.33 | 0.33 | 0.67 | 0.67 |
| Trigram | 0.00 | 0.00 | 0.50 | 0.50 |
| 4-gram | 0.00 | 0.00 | 0.32 | 0.32 |
| Skip-bigram | 0.38 | 37.5 | 0.64 | 0.64 |
| LCS | 0.53 | 0.53 | 0.73 | 0.73 |
| Synonyms | NA | 0.85 | NA | 0.88 |
| Relationship | NA | 0.85 | NA | 0.88 |
| Unigram (POS) | 0.85 | 0.85 | 0.88 | 0.88 |

Example 3 is a representation of changing the sentence structure/order. As the sequence of the words in the reference sentence had been changed in the candidate sentence, LCS was obviously affected by this type of plagiarism. Because usually LCS had similar scores like unigram but in this case its score was significantly lower than unigram. Actually bigram to 4-gram were also affected, especially after the stopwords were removed. This was probably because the order of open class words had been changed forming different n-grams.

Example 4:

Candidate Sentence: the **increase** of industry the growth of cities and the **explosion** of the population were three **large** factors of nineteenth century america

Reference Sentence: the **rise** of industry the growth of cities and the **expansion** of the population were the three **great** developments of late nineteenth century american history

TABLE 8 Scores of Example 4

| Methods | SW+SM | SW | SM | No Pre |
|---|---|---|---|---|
| Unigram | 0.50 | 0.50 | 0.72 | 0.72 |
| Bigram | 0.27 | 0.27 | 0.49 | 0.49 |
| Trigram | 0.10 | 0.10 | 0.37 | 0.37 |
| 4-gram | 0.00 | 0.00 | 0.29 | 0.29 |
| Skip-bigram | 0.20 | 0.20 | 0.55 | 0.55 |
| LCS | 0.50 | 0.50 | 0.72 | 0.72 |
| Synonyms | NA | 0.67 | NA | 0.81 |
| Relationship | NA | 0.82 | NA | 0.89 |
| Unigram (POS) | 0.50 | 0.50 | 0.72 | 0.72 |

This example demonstrated how original content could be modified with synonyms. Totally different words were used leading to false negative judgment on the similarity between two sentences. The impact was more obvious after stopwords were removed. However, the WordNet-based methods performed pretty well in such situation, relationship-based method in particular.

The strengths and weaknesses of the methods were determined through the analysis of the above four examples. Meaningful observations were made to learn about the characteristics of each method and assess the value of each method in plagiarism detection. One of the observations was that stemming did not have obvious influence on the detection results. This argument was derived from the statistics of the abstract set and was further confirmed by the paraphrased set. The cause of this phenomenon is probably due to the scope of detection in this research; the probability of words "computer", "computation", and "compute" to appear in a pair of sentences can be reasonably assumed to be lower than the probability of the same words to appear in a pair of documents, limiting the effect of stemming. The observations also raised some concerns about a couple of probable scenarios, which have not been discussed in this research. The scenarios expose the weaknesses of the proposed methods and they are as follows:

1. Scenario 1 – a word is substituted with a phrase bearing the same meaning. For example, "regardless" is substituted with "no matter". In this case, "no matter" will be treated as two individual words, so even the WordNet-based methods cannot detect the substitution.

2. Scenario 2 – two or more original sentences are combined into a long sentence, or an original sentence is split into two shorter sentences. As all proposed methods adopt sentence to sentence comparison, similarity between reference sentence and candidate sentence will be greatly affected because the number of matched words is divided by a larger denominator (long candidate sentence), or the number of matched words decreases (divide into separate sentences).

## 5. Conclusion

This study proposed implementation of ROUGE, which was previously applied in the evaluation of text summarization, to detect plagiarism. WordNet was introduced to integrate with ROUGE in order to handle as many forms of plagiarism as possible.

Through the analysis of the experimental results, the proposed methods are proven to have research value in the field of plagiarism detection. Each method has its strengths in dealing with certain types of plagiarism; while at the same time, each has its weaknesses in certain situations. ROUGE is capable of detecting verbatim copy, and the efficiency is acceptable when comparing two complete documents. Unigram is not bounded by the in-sequence constraint like LCS and other n-grams. Other n-grams are stricter in matching tokens and therefore they have higher precision. While LCS and skip-bigram take a middle ground because both allow skips when scanning through a sentence but require matching tokens to be in-sequence. WordNet looks into the semantic aspect of words so that matching is not just by exact match, but also by the meanings of the tokens.

In the future, several tasks will be conducted. First, fur-ther confirmation on the effectiveness of the proposed methods and recommended settings can be achieved by running tests with a larger and more diversified corpus. Second, the proposed methods were tested and evaluated separately in this research. To combine different methods together, a weighting scheme should be developed so that the score of each method contributes in the right proportion and the final score at the end accurately represent the methods involved. Third, to overcome the problem of splitting or integrating original sentence(s) into one or more sentence(s), chunk comparison may be a worthy attempt. The inclusion of neighboring sentences and comparison of these sentences as chunks should be able to solve this loophole. Application of n-gram to chunk comparison can make the method more robust. New chunks can be formed with each in-sequence consecutive sentence. Last but not least, hopefully the system can be further developed for educational purpose by adding educative messages and explanations regarding the detection results. By providing explanations according to the types of plagiarism, users (including students) can better understand plagiarism and know how to avoid it with real examples.

## Acknowledgment

This work was supported in part by a grant from National Science Council with grant no. NSC96-2520-S-009-003-MY3.

**Chien-Ying Chen** received his master degree in 2008 from Institute of Information Management, National Chiao Tung University, Taiwan, R.O.C. His main research interests are in information retrieval, Web mining, and plagiarism detection.

**Jen-Yuan Yeh** received his master degree in 2002 and Ph.D. degree in 2008, both in Computer and Information Science, from National Chiao Tung University, Taiwan, R.O.C. He currently works for the National Museum of Natural Science. His main research interests are in text summarization, digital archives, information retrieval, and Web mining.

**Hao-Ren Ke** received his B.S. degree in 1989 and Ph.D. degree in 1993, both in Computer and Information Science, from National Chiao Tung University, Taiwan, R.O.C. He is currently a professor of Graduate Institute of Library and Information Studies, National Taiwan Normal University and the Deputy Library Director. His main research interests are in digital library/archives, library and information system, information retrieval, and Web mining.